\documentclass[twocolumn,twocolappendix]{aastex631}

\usepackage{ps_macros}

\defcitealias{bunkerJADESNIRSpecSpectroscopy2023}{B23}
\defcitealias{cameronNitrogenEnhancements4402023}{C23}

\shorttitle{}
\shortauthors{}

\graphicspath{{./}{figs/}}

\newcommand{\gnz}{GN-z11}
\newcommand{\mrk}{Mrk~996}

\newcommand{\Uav}{\hbox{$\langle U\rangle$}}

\begin{document}

\title{\gnz{} in context: possible signatures of globular cluster precursors at redshift 10}

\author[0000-0002-9132-6561]{Peter Senchyna}
\altaffiliation{Carnegie Fellow}
\email{psenchyna@carnegiescience.edu}
\affiliation{The Observatories of the Carnegie Institution for Science, 813 Santa Barbara Street, Pasadena, CA 91101, USA}

\author[0000-0003-0390-0656]{Adele Plat}
\affiliation{Steward Observatory, University of Arizona, 933 N Cherry Ave, Tucson, AZ 85721, USA}

\author{Daniel P.\ Stark}
\affiliation{Steward Observatory, University of Arizona, 933 N Cherry Ave, Tucson, AZ 85721, USA}

\author{Gwen C. Rudie}
\affiliation{The Observatories of the Carnegie Institution for Science, 813 Santa Barbara Street, Pasadena, CA 91101, USA}

\begin{abstract} 
\noindent
The first \jwst{} spectroscopy of the luminous galaxy \gnz{} simultaneously both established its redshift at $z=10.6$ and revealed a rest-ultraviolet spectrum dominated by signatures of highly-ionized nitrogen, which has so far defied clear interpretation.
Here we present a reappraisal of this spectrum in the context of both detailed nebular modeling and nearby metal-poor reference galaxies.
The \ion{N}{4}] emission enables the first nebular density measurement in a star-forming galaxy at $z>10$, and reveals evidence for extremely high densities $n_e\gtrsim 10^5$~$\mathrm{cm^{-3}}$.
We definitively establish with a suite of photoionization models that regardless of ionization mechanism and accounting for depletion and this density enhancement, an ISM substantially enriched in nitrogen ($[\mathrm{N/O}]=+0.52$) is required to reproduce the observed lines.
A search of local UV databases confirms that nearby metal-poor galaxies power \ion{N}{4}] emission, but that this emission is uniformly associated with lower densities than implied in \gnz{}.
We compare to a unique nearby galaxy, Mrk~996, where a high concentration of Wolf-Rayet stars and their CNO-processed wind ejecta produce a UV spectrum remarkably similar to that of both \gnz{} and the Lyc-leaking super star cluster in the Sunburst Arc.
Collating this evidence in the context of Galactic stellar abundances, we suggest that the peculiar nitrogenic features prominent in \gnz{} may be a unique signature of intense and densely clustered star formation in the evolutionary chain of the present-day globular clusters, consistent with in-situ early enrichment with nuclear-processed stellar ejecta on a massive scale.
Combined with insight from local galaxies, these and future \jwst{} data open a powerful new window onto the physical conditions of star formation and chemical enrichment at the highest redshifts.
\end{abstract}

\keywords{High-redshift galaxies (734) --- Blue compact dwarf galaxies (165) --- Galaxy abundances (574) --- Globular star clusters (656)}

\section{Introduction}
\label{sec:intro}

With \jwst{} operational, some of the earliest epochs of star formation are now within reach.
Many questions remain about these early times, from the properties and distribution of the first unenriched stellar populations to the character of stellar mass assembly across different high-redshift halos \citep[e.g.][for reviews]{brommFirstStars2004,brommFirstGalaxies2011,robertsonGalaxyFormationReionization2022}.
One of the most clear predictions about this era is that many galaxies must have been host to bursts of extremely intense, highly-clustered star formation.
The evidence for and antecedents of at least some of these episodes are the large populations of globular clusters in the present-day Universe, whose typical metallicities ($[\mathrm{Fe/H}]\sim -2.5$--$-1$; \citealt{harrisCatalogParametersGlobular1996}) and inferred ages (uncertain, but ranging from $\sim 10$ to $\gtrsim 13$~Gyr; \citealt{vandenbergAgeGalacticGlobular1996,brodieExtragalacticGlobularClusters2006}) establish many as potentially forming at $z\gtrsim 5$ \citep[e.g.][]{renziniFindingFormingGlobular2017}.

Detailed study of globulars has also provided a mysterious clue about the character of the star formation episodes that produced them.
Essentially all globulars differ from a single simple stellar population: and in particular, a substantial fraction of their stars (their `enriched' or `second` population) contain variations in light and heavy element abundances consistent with the yields of high-temperature nuclear burning (specifically, the CNO process and proton-capture chains, which produce abundance N, Na, and Al) that sharply distinguish them from field star populations \citep[e.g.][]{osbornTwoNewCNstrong1971,smithChemicalInhomegeneityGlobular1987,grattonAbundanceVariationsGlobular2004,carrettaAbundancesSlightlyEvolved2005,carrettaDetailedAbundancesLarge2010}.
However, the exact origins of these abundance patterns remains unclear; and even the basic question of whether these patterns were inherited from earlier generations of stars or are an evolutionary effect remains unsettled \citep[e.g. discussion in][]{bastianMultipleStellarPopulations2018}.

The great promise of \jwst{} lies in its ability to begin to inform mysteries about the early Universe like these by directly probing high-redshift star formation in-situ.
Deep \hst{} imaging had already unveiled extremely compact star-forming clumps at $z\sim 2--10$ via gravitational lensing with star formation rate surface densities and radii approaching expectations for proto-globular clusters \citep[e.g.][]{rigbyStarFormation4812017,vanzellaPavingWayJWST2017,vanzellaMUSEDeepLensed2021b,welchRELICSSmallscaleStar2023}.
However, these clumps were typically too faint even with lensing for spectroscopic study pre-\jwst{}, with the exception of the most massive and lowest-redshift lensed examples \citep[e.g.][]{pascaleLymancontinuumleakingSuperStar2023,mestricCluesPresenceSegregation2023}.
\jwst{} imaging alone has already begun revolutionizing our understanding of this area, with some of the first images revealing surprisingly dense clusters in lensed fields \citep[e.g.][]{claeyssensStarFormationSmallest2023} and bright $z\sim 6-8$ galaxies uniformly resolving into small star-forming clumps even in the rest-optical \citep{chenJWSTNIRCamObservations2023}.

The first \jwst{} rest-UV and optical spectroscopy at these highest redshifts has already provided substantial insight into the physical conditions in the earliest star-forming galaxies.
These data have confirmed that many $z\sim 6$--$8$ systems are undergoing intense bursts of star formation, and revealed evidence for moderately-low metallicities that suggest a fairly rapid build-up of metals in these systems \citep[e.g.][]{arellano-cordovaFirstLookAbundance2022,schaererFirstLookJWST2022,trumpPhysicalConditionsEmissionLine2022,jonesEarlyResultsGLASSJWST2023,tangJWSTNIRSpecSpectroscopy2023}.
The rest-frame UV is a particularly powerful portion of these spectra, providing insight onto massive stars and highly-ionized gas in these systems.
With a few uncertain exceptions, the rest-UV spectra obtained thus far are reasonably close in agreement to those of `typical' metal-poor star-forming galaxies in the local Universe, where \ion{C}{3}], \ion{O}{3}], \ion{He}{2}, and \ion{C}{4} are the most prominent transitions \citep[e.g.][]{senchynaUltravioletSpectraExtreme2017,senchynaExtremelyMetalpoorGalaxies2019,bergIntenseIVHe2019,mingozziCLASSYIVExploring2022}.

In contrast, \jwst{} observations of \gnz{} represent a significant deviation from expectations.
\gnz{} was the first spectroscopically-confirmed galaxy at $z>10$ \citep{oeschRemarkablyLuminousGalaxy2016}.
Its confident early identification as a $J_{125}$-band Lyman-break dropout candidate in \hst{}/WFC3/IR imaging \citep{oeschMostLuminousTextasciitilde2014} was due to its remarkable UV luminosity: $M_\mathrm{UV} = -22$, roughly a magnitude ($3\times$) brighter than the characteristic luminosity derived at $z\sim 7$--8.
A range of observations of \gnz{} were conducted as part of the \jwst{} Advanced Deep Extragalactic Survey (JADES) in February 2023.
Results from deep imaging with NIRCam were described by \citet{tacchellaJADESImagingGNz112023}, revealing a compact morphology dominated by a $\lesssim 200$~pc half-light radius core with at least half the total light in an unresolved point-source component.
Population synthesis modeling fits to the \jwst{} photometry indicate that this core harbors a $\sim 10^{8-9}$~$\mathrm{M_\odot}$ stellar population formed within the last $\sim 10$--$20$~Myr, with a current SFR of $\sim20$~$\mathrm{M_\odot/yr}$ and corresponding specific star formation rate (sSFR, SFR/$M_\star$) of $\sim 20$--30~$\mathrm{Gyr^{-1}}$.
This dominant clumpiness is in-keeping with the properties of other luminous $z\gtrsim 6$ galaxies observed with \jwst{} \citep[e.g.][]{chenJWSTNIRCamObservations2023}; but for such a particularly luminous source to resolve into a single nearly-unresolved core is surprising.
The star formation rate surface density implied by these obsevations is $\Sigma_\mathrm{SFR} \gtrsim 40$~$\mathrm{M_\odot/yr/kpc^2}$ --- a range reached only by the most intense blue compact dwarfs and ultraluminous infrared galaxies in the local Univeres \citep[e.g.][]{kennicuttStarFormationMilky2012}.

The \jwst{}/NIRSpec spectra reported by \citet[][hereafter \citetalias{bunkerJADESNIRSpecSpectroscopy2023}]{bunkerJADESNIRSpecSpectroscopy2023} are yet more surprising.
The rest-UV spectrum of \gnz{} is dominated by emission lines, with the most prominent being \ion{N}{4}] $\lambda 1486$ and \ion{N}{3} $\lambda 1750$, followed by \ion{C}{3}] $\lambda\lambda 1907,1909$.
This is in stark contrast to star-forming galaxy samples at lower redshifts, where these transitions are rarely detected and essentially always at substantially lower flux than the \ion{O}{3}] doublet \citep[e.g.][]{amorinAnaloguesPrimevalGalaxies2017,mingozziCLASSYIVExploring2022,saxenaStrongIVEmission2022}.
In a follow-up analysis, \citet[][or \citetalias{cameronNitrogenEnhancements4402023}]{cameronNitrogenEnhancements4402023} suggest this spectrum requires extremely elevated $\mathrm{N/O}>-0.5$, and discuss scenarios including runaway stellar collisions producing extremely massive stars or tidal disruption events that might be capable of reproducing the spectral features.

In this work, we present a reappraisal of this striking \jwst{} spectrum in the context of nebular models and an extensive search of UV spectra of nearby galaxies.
In Section~\ref{sec:niv}, we reconsider the gas conditions implied by the \ion{N}{4}] detection in-particular.
We proceed in Section~\ref{sec:photomod} to analyze the \gnz{} UV--optical nebular line spectrum in the context of a detailed suite of photoionization models.
In Section~\ref{sec:localref}, we seek out lower-redshift star-forming galaxies with UV spectra close to that observed in \gnz{}.
And in Section~\ref{sec:summary}, we combine the inferences from the previous section into a picture of \gnz{} that we consider in a broader context.

We assume the Planck 2018 cosmological parameters throughout \citep{collaborationPlanck2018Results2020}; and adopt the solar abundance patterns compiled by \citet{bressanPARSECStellarTracks2012}, primarily relying on \citet{caffauSolarChemicalAbundances2011} but modified in the photoionization modeling according to \citet{gutkinModellingNebularEmission2016}.

\section{Constraints on nebular density from \ion{N}{4}] emission}
\label{sec:niv}

The detection of \ion{N}{4}] visible in the NIRSpec grating spectrum of \gnz{} readily provides a clue about the state of the highly-ionized gas in the galaxy.
The \ion{N}{4} ion is a Be-like $2s^2$ species analogous to \ion{C}{3}; and \ion{N}{4}] $\lambda 1486$ is part of a forbidden/semi-forbidden doublet representing the same pair of ground state electronic transitions as [\ion{C}{3}]~$\lambda 1907$, \ion{C}{3}]~$\lambda 1909$ \citep[e.g.][]{osterbrockAstrophysicsGaseousNebulae2006}.
The forbidden magnetic quadrupole transition counterpart to the intercombination electric dipole 1486~\AA{} transition is [\ion{N}{4}] $\lambda 1483$. 
As with the \ion{C}{3} doublet, the ratio of the two components is sensitive to the electron density $n_e$; and both have relatively high critical densities ($5 \times 10^5$~$\mathrm{cm}^{-3}$ for \ion{C}{3} and $1\times 10^6$~$\mathrm{cm}^{-3}$ for \ion{N}{4}).

\begin{figure*}
    \centering
    \includegraphics[width=\textwidth]{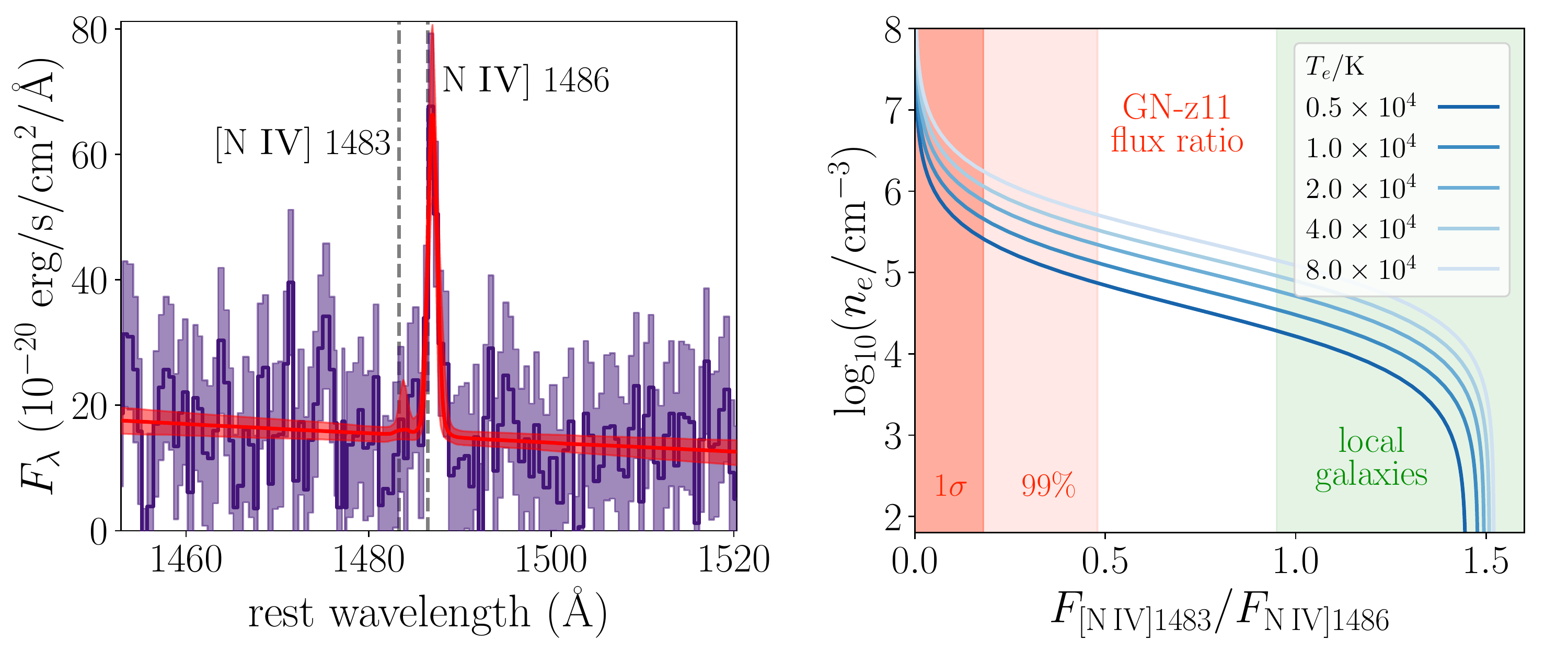}
    \caption{
        \ion{N}{4}] emission detected in \gnz{} provides evidence for a very high electron density in the emitting gas. 
        Fits to the \jwst{}/NIRSpec grating spectrum place a strong upper limit on the flux of the forbidden [\ion{N}{4}]~$\lambda 1483$ counterpart to \ion{N}{4}]~$\lambda 1486$ (left).
        Such low inferred flux ratios are indicative of efficient collisional de-excitation of the 1483 transition, achieved at extremely high electron densities $n_e \gtrsim 10^5$-$10^6$~$\mathrm{cm}^{-3}$ (right).
        These high densities are rarely found in star-forming galaxies; for comparison, the range of \ion{N}{4}] flux ratios we measure for a set of local metal-poor blue compact dwarfs (Section~\ref{sec:localref}) are plotted in green, and found to be consistent with $n_e\sim 10^4$--$10^5$~$\mathrm{cm^{-3}}$ or lower.
    }
    \label{fig:niv_ratio}
\end{figure*}

Intriguingly, while \ion{N}{4}]~$\lambda 1486$ is strongly detected in \gnz{}, no emission is evident in [\ion{N}{4}]~$\lambda 1483$.
The \jwst{}/NIRSpec grating spectrum clearly resolves the two transitions at their expected wavelength separation at the redshift of \gnz{}.
To quantify the constraint the \jwst{} data place on the flux ratio, we analyzed a manually-digitized\footnote{Using \texttt{WebPlotDigitizer} v4.6 (\url{automeris.io/WebPlotDigitizer})} version of the grating spectrum as presented in \citetalias[][their Figure~2]{bunkerJADESNIRSpecSpectroscopy2023}.
We then fit a pair of Gaussians with flexible but tied widths and wavelength ratio fixed to the expected separation of the two transitions \citep{kramidaNISTAtmoicSpectra2022}, and repeat this procedure 500 times while bootstrap-resampling according to the conservative approximated flux uncertainty vector (we assume this uncertainty vector is constant-valued, and scale it to visually match the uncertainty plotted in the original figure).
This results in a $1\sigma$ upper-limit on the $F_{1483}/F_{1486}$ of $<0.18$; and conservatively at the $99^{\mathrm{th}}$-percentile level $F_{1483}/F_{1486}<0.48$ (Figure~\ref{fig:niv_ratio}).

To interpret this ratio in terms of the nebular physical conditions, we use the code \texttt{PyNeb} \citep{luridianaPyNebNewTool2015} to compute a grid of line emissivities\footnote{With transition probabilities and collision strengths for \ion{N}{4} from \citet{wieseAtomicTransitionProbabilities1996} and \citet{ramsbottomElectronImpactExcitation1994}, respectively} at varying electron temperature $T_e$ and density $n_e$.
We find that as-expected, the ratio of the forbidden to semi-forbidden component $F_{1483}/F_{1486}$ falls with increasing electron density and, to a lesser degree, with decreasing electron temperature.
The resulting predicted line ratios are shown in Figure~\ref{fig:niv_ratio} (right panel), in comparison to the line ratio constraints derived from the \jwst{} spectrum.
Even assuming a conservatively-low $T_e = 0.5\e{4}$~K, the 99\%-confidence upper-limit on the $F_{1483}/F_{1486}$ flux ratio implies $n_e \gtrsim 10^5$--$10^6$~$\mathrm{cm^{-3}}$ in the \ion{N}{4}-emitting gas.

As noted above, the analogous \ion{C}{3}] doublet provides a complementary probe of density in lower-ionization gas (ionization potential of 24.4 versus 47.4 eV).
Unfortunately, \ion{C}{3}] is not cleanly separated by the medium-resolution NIRSpec gratings used by JADES in the \gnz{} observations; but the profile is resolved and clearly broader than the other nebular lines.
We apply the same technique described above to fit the \ion{C}{3}] doublet, assuming a similar linewidth for each component to \ion{N}{4}]. 
Consistent with the qualitatively symmetric profile shape, we infer a $1\sigma$ confidence interval on the line ratio of $F_{1907}/F_{1909}= 1.0_{-0.3}^{+0.4}$.
While not strongly constraining, a similar comparison with \texttt{PyNeb} indicates that this confidence interval on the line ratio is consistent with a lower density ($n_e\simeq 10^4$~$\mathrm{cm^{-3}}$).
We briefly note that the \ion{N}{3}] complex is also a density probe \citep[e.g.][]{keenanIIIIntercombinationLines1994}; but it is not strongly sensitive over the considered range, and the profile in \gnz{} is noisy (Figure~\ref{fig:mrk996comp}).
If the \ion{N}{4} and \ion{C}{3} emission originate in the same \hii{} regions, our \ion{C}{3}] $n_e$ estimate is suggestive of a significant density gradient between the highly ionized regions close-in to the ionizing sources where \ion{N}{4}] is produced and the further-out, lower-ionization \ion{C}{3}]-emitting gas.

The constraints we have derived here leveraging the \ion{N}{4}] doublet represent the first direct inference on density in a star-forming galaxy at $z>10$, and reveal an extraordinarily high electron density $n_e\gtrsim 10^5$~$\mathrm{cm^{-3}}$.
Previous studies investigating the evolution of electron density with redshift have found notable increases from $z\sim 0$ to 2--3 and recently with \jwst{} out to $z\sim 9$ \citep[e.g.][]{brinchmannNewInsightsStellar2008,hainlineRestFrameOpticalSpectra2009,sandersMOSDEFSurveyElectron2016,isobeRedshiftEvolutionElectron2023}, but over a disjoint range of densities $n_e \simeq 10^2$--$10^3~\mathrm{cm^{-3}}$ probed by the lower-ionization and lower-$n_{\mathrm{crit}}$ optical [\ion{O}{2}] and [\ion{S}{2}] diagnostics.
Work with the \ion{C}{3}] doublet and other higher-ionization lines like \ion{Ar}{4} have clearly established that in individual star-forming systems both locally and further afield, these lines frequently evince the presence in the same galaxies of much higher density gas (often by orders of magnitude) than implied by lower-$n_{\mathrm{crit}}$ tracers \citep[e.g.][]{jamesMappingUVProperties2018,mingozziCLASSYIVExploring2022,mainaliSpectroscopyCASSOWARYGravitationallylensed2023}.
But even in this context, densities as high as are implied by the limit we place on the \ion{N}{4}] ratio in \gnz{} are exceptionally rare in star-forming galaxies.
The \ion{C}{3}] doublet and even \ion{N}{4}] where it is detected at lower redshifts (see Figure~\ref{fig:niv_ratio} and Section~\ref{sec:localref}) are essentially always consistent with lower densities of $10^4$--$10^5~\mathrm{cm^{-3}}$ at most.
Our measurement in \gnz{} is suggestive of particularly dense conditions in the innermost highly-ionized zones of the dominant \hii{} regions; such high densities are more generally associated with environments such as AGN, ultracompact \hii{} regions, and nebulae around symbiotic stars composed of dense stellar ejecta \citep[e.g.][]{kenyonNatureSymbioticStars1984,woodMorphologiesPhysicalProperties1989,veron-cettyEmissionLineSpectrum2000}.
In the following sections we explore the possible implications of this high inferred density and whether analagous processes might help explain their occurrence in this luminous $z=10.6$ galaxy.

\section{Reproducing \gnz{}-like line ratios with photionization models}
\label{sec:photomod}

We next turn our attention to understanding the anomalously-strong flux of the nitrogen lines relative to the other rest-UV lines in \gnz{}.
These fluxes have already been analyzed by \citetalias{cameronNitrogenEnhancements4402023}, who use a simple two-zone nebula model to argue that the \ion{N}{3}]/\ion{O}{3}] ratio implies highly elevated nitrogen relative to oxygen: conservatively concluding N/O$>-0.49$.
And \citetalias{bunkerJADESNIRSpecSpectroscopy2023} leverage a UV line ratio diagram to suggest consistency with a stellar ionizing spectrum, but do not consider whether such models reproduce the nitrogen lines.
These analyses represent a critical first step, but are limited by several uncertainties acknowledged by the authors.
First, they do not quantitatively consider the flux in the \ion{N}{4}] line and whether it is consistent with stellar photoionization or implies an even higher N/O abundance.
And second, they do not consider in-detail the effects of depletion, extremely high gas densities (Section~\ref{sec:niv}), and other ionization mechanisms, all of which might affect the nitrogen line ratios and provide an alternative explanation avoiding such elevated N/O.

To investigate these questions, in this section we consider the full set of nebular line strengths in \gnz{} in the context of a suite of detailed photoionization models which we describe here.
The primary grid we rely upon is computed with \texttt{CLOUDY} c17.02 \citep{ferland2017ReleaseCloudy2017} following the method outlined in detail by \citet{gutkinModellingNebularEmission2016,platConstraintsProductionEscape2019} and summarized here.
It employs the latest predictions of the Charlot \& Bruzual stellar population synthesis models which include updated prescriptions for the evolution and atmospheres of massive stars as ionizing sources \citep[see also discussion in][]{senchynaUltravioletSpectraExtreme2021,senchynaDirectConstraintsExtremely2022}.
We vary key parameters including the the zero-age volume-average ionization parameter \Uav; the gas density $n_H\sim n_e$; as well as $\mathrm{N/O}$.
We then produce tracks of line ratios over varying stellar population ages at a range of total metallicities.
The fiducial model follows eq.11 of \citet{gutkinModellingNebularEmission2016} for the abundances of nitrogen and oxygen; which predicts that N/O increases with oxygen abundance to account for primary and secondary nucleosynthetic components, and reproduces the relatively tight relationship between these abundances found in local  star-forming galaxies, \hii{} regions, and in the field star population of the Milky Way \citep[e.g.][]{bergDirectOxygenAbundances2012,israelianGalacticEvolutionNitrogen2004}.
This relationship predicts $(\mathrm{N/O})_{\mathrm{tot}}=-1.8$ at $\log_{10}(Z/Z_\odot)=-0.94$ ($12+\log_{10}(\mathrm{O/H})_{\mathrm{gas}}=7.8$). 

The impact of dust depletion is particularly crucial in estimating N/O.
Since O is refractory and N is not, significant depletion onto dust grains could artificially boost N/O in the gas-phase and mimic a genuine nitrogen enhancement.
The \texttt{CLOUDY} grid we run incorporates a self-consistent treatment for this effect \citep[see][]{gutkinModellingNebularEmission2016}, parameterized by the dust-to-metal mass ratio $\xi_d$.
We consider values of $\xi_d$ ranging from 0.1 to 0.3 \citep[near the solar value;][]{gutkinModellingNebularEmission2016} up to extremely high values of 0.95 below.

In addition to ionization by stellar populations, we consider model predictions for both narrow-line AGN and fast radiative shocks.
For AGN, we compare to the models produced by \citet{feltreNuclearActivityStar2016} with $-1.2<\log_{10}(Z/Z_\odot)<0.2$, $0.1<\xi_d<0.5$, $-4\leq\log_{10}(\Uav)\leq-1$, and a power-law index varying from $-2$ to $-1.2$. 
For shocks, we use the radiative shock models of \citet{alarieExtensiveOnlineShock2019}, including both the shock and precursor components. 
We consider shock velocities varying from $10^{2}$ to $10^{3}\, \rm{km} \rm{s}^{-1}$, a pre-shock density of $10^{2} \, \rm{cm}^{-3}$, a transverse magnetic field of 1 $\mu G$, and ISM metallicities $-1.2<\log_{10}(Z/Z_\odot)<0.2$. 
Both of these models follow the same nominal assumed relation between N/O and O/H as described above.

\begin{figure*}
    \centering
    \includegraphics[width=\textwidth]{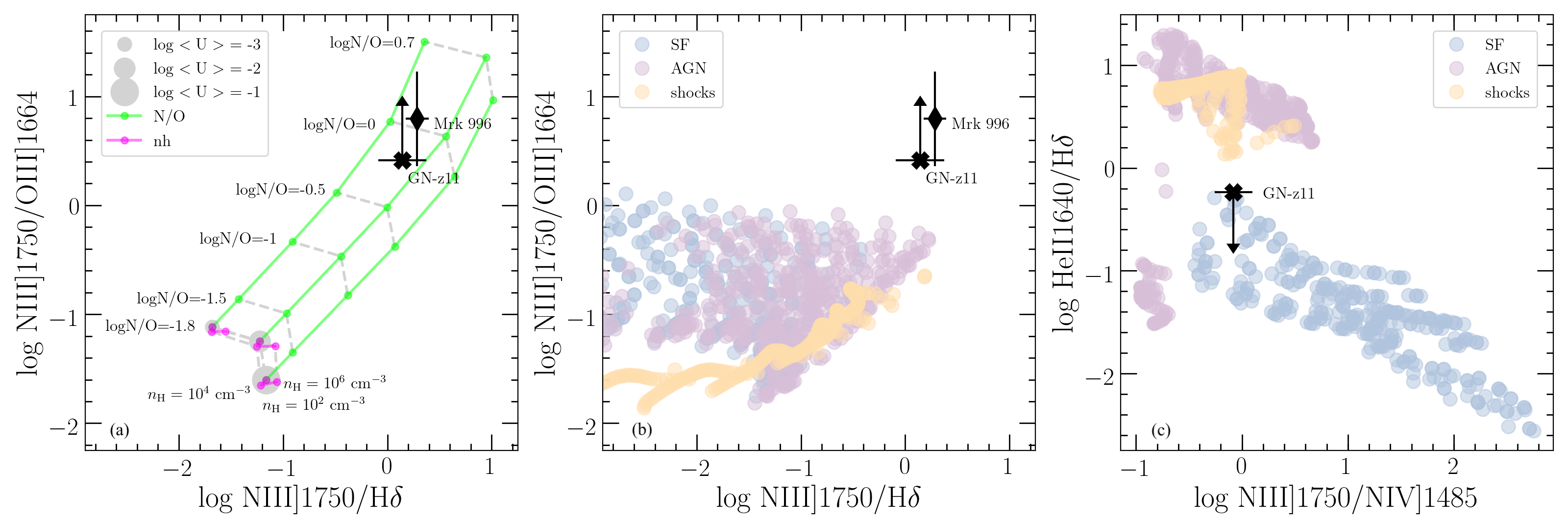}
    \caption{Comparison of \gnz{} (cross) and \mrk{} (diamond) line ratios with predictions from photoionization models. (a) $\log_{10}$(\ion{N}{3}]~$\lambda 1750$/\ion{O}{3}]~$\lambda 1664$) versus $\log_{10}$(\ion{N}{3}]~$\lambda 1750$/H$\delta$).
    Star-forming galaxy models are shown for a metallicity log(Z/Z$_\odot$)=-0.94, $\xi_d$=0.3 for log(U)=-1, -2 and -3.The green lines show the increase of $\log_{10}$(N/O) from -1.8 to 0.7 and the fuchsia lines show the increase of hydrogen density from $10^{2} \rm{cm}^{-3}$ to $10^{6} \rm{cm}^{-3}$. (b) $\log_{10}$(\ion{N}{3}]~$\lambda 1750$/\ion{O}{3}]~$\lambda 1664$) versus $\log_{10}$(\ion{N}{3}]~$\lambda 1750$/H$\delta$). The circles show the predictions from photoionization models for star-forming galaxy models (blue), AGN models of \citet{feltreNuclearActivityStar2016} (purple) and radiative shock models of \citet{alarieExtensiveOnlineShock2019} (yellow). (c) Same as diagram (b) but for $\log_{10}$(\ion{N}{3}]~$\lambda 1750$/\ion{N}{4}]~$\lambda 1485$) versus $\log_{10}$(\ion{He}{2}~$\lambda 1640$/H$\delta$).
    The observed line ratios are in best agreement with photoionization by a metal-poor stellar ionizing spectrum; but regardless of ionization source, they conclusively require a significant enhancement in N/O in the ionized gas.
    }
    \label{fig:photomod}
\end{figure*}

In Figure~\ref{fig:photomod}, we compare key measured line ratios in \gnz{} to these model predictions. 
First, we compare all three grids in panel (b) of Figure~\ref{fig:photomod} to the observed constraints on \ion{N}{3}]/\ion{O}{3}] versus \ion{N}{3}]/H$\delta$.
The star-forming models plotted span $-1.2<\log_{10}(Z/Z_\odot)<0.2$ with $Z_\star=Z_{\mathrm{ISM}}=Z$, $0.1<\xi_d<0.5$, and $-4\leq\log_{10}(\Uav)\leq-1$ for a 10~Myr constant star-formation. 
The star-forming models with nominal N/O are clearly offset from the observed line ratio in \gnz{}, as argued by \citetalias{cameronNitrogenEnhancements4402023}.
The radiative shocks and AGN models are capable of powering significantly-higher \ion{N}{3}]/H$\delta$ than star-forming galaxy models, reaching closer to the observed line ratios without invoking an enhancement in nitrogen abundance.
This reflects the impact of the more highly-ionized and hotter gas these models are capable of producing.
However, both AGN and shocks with nominal N/O fail to reproduce the large \ion{N}{3}]/\ion{O}{3}] ratio of \gnz{}, as well as its large \ion{N}{3}]/\ion{C}{3}] ratio.

Next we consider the \ion{N}{3}]/\ion{N}{4}] ratio and \ion{He}{2}/H$\delta$ (Figure~\ref{fig:photomod}), both of which act as proxies for ionization.
The \ion{He}{2}/H$\delta$ recombination line ratio is effectively a direct probe of the ionizing power at $54.4$ versus $13.6$~eV assuming no stellar contributions \citep[e.g.][]{senchynaExtremelyMetalpoorGalaxies2019}; while the \ion{N}{3}]/\ion{N}{4}] ratio probes ionization conditions sensitive to the relative flux at the slightly lower ionization potential of \ion{N}{3} (47.4~eV).
While most of the stellar models prefer stronger \ion{N}{3}] than \ion{N}{4}], the near-unity ratio found in \gnz{} can be reproduced at high ionization parameters $\log_{10}(\Uav)>-1.7$.
In addition, models with lower stellar metallicities also prefer lower \ion{N}{3}]/\ion{N}{4}] reflecting their harder ionizing spectra.
We compute additional models with the stellar metallicity decoupled from the ISM metallicity \citep[simulating expected enhancements in $\alpha$/Fe; e.g.][]{steidelReconcilingStellarNebular2016,stromMeasuringPhysicalConditions2018,senchynaDirectConstraintsExtremely2022} and find that these produce even lower ratios.
Even without invoking this effect, the implied ionization parameter is not unreasonable.
In a spherical geometry, $\Uav \propto (Q_{\rm{H}}\epsilon^2n_\mathrm{H})^{1/3}$ \citep[e.g.,][]{charlotNebularEmissionStarforming2001}. 
Since \Uav\ is proportional to $n_{\mathrm{H}}^{1/3}$, the high density conditions we infer from the \ion{N}{4}] doublet may itself help explain why it is so prominent relative to \ion{N}{3}].

These low values of \ion{N}{3}]/\ion{N}{4}] are also consistent with the ionization conditions produced by AGN and shocks.
However, the upper limit on \ion{He}{2} is constraining in this case, as pointed out by \citetalias{bunkerJADESNIRSpecSpectroscopy2023}.
Though it is compatible with the values predicted by the stellar models and even the discrepantly-strong \ion{He}{2} emission routinely observed in local metal-poor star-forming galaxies \citep[e.g.][]{shiraziStronglyStarForming2012,senchynaHighmassXrayBinaries2020,bergCharacterizingExtremeEmissionline2021}, it is lower than the extremely high values predicted by the AGN and shock ionized models at the same \ion{N}{3}]/\ion{N}{4}].
Likewise, while less discriminating \citep[see e.g.][]{feltreNuclearActivityStar2016}, the lack of \ion{N}{5} emission (ionization potential 77.5~eV) in the \jwst{} spectrum argues against the much harder ionizing spectra these source produce.
While we cannot definitively rule-out a contribution from AGN or shocks, we find better consistency with models dominated by star formation.

As we have seen, however, none of the models reproduce \ion{N}{3}]/(\ion{O}{3}], \ion{C}{3}]) at nominal N/O.
We explore several of the other potential variables impacting this ratio in the first panel of Figure~\ref{fig:photomod}.
The starting point displayed is a 4~Myr constant star formation history model with a stellar and ISM metallicity $\log_{10}(Z/Z_\odot)=-0.94$; we choose this value for Z as this is the best fit inferred for \gnz{} (see below). 
We adopt a nominal $\xi_d=0.3$ and plot ionization parameters $\log_{10}$(\Uav)=-1, -2 and -3. 

First, we explore the impact of increasing the gas density on the UV--optical line emission.
The fiducial model has $n_{\rm{H}}=10^{2}\, \rm{cm}^{-3}$, and we consider values increased up to the range inferred from the \ion{N}{4}] doublet ratio (Section~\ref{sec:niv}).
Increasing the density even up to $10^6 \, \mathrm{cm^{-3}}$ does not strongly change these line ratios (Figure~\ref{fig:photomod}).
The main effect of increasing the density above $10^4 \, \mathrm{cm^{-3}}$ is the decrease of [\ion{O}{2}]~$\lambda 3727$ compared to the other lines since, for an electronic temperature of $10^4$~K, the critical density for collisional deexcitation of \ion{O}{2} $^2D_{5/2}^0$ and $^2D_{3/2}^0$ are $3.4 \times 10^3$ and $1.5 \times 10^4$ respectively \citep[][]{osterbrockAstrophysicsGaseousNebulae2006}. 
These models with high density ($10^6 \, \mathrm{cm^{-3}}$) underpredict [\ion{O}{2}]/H$\delta$ and [\ion{O}{2}]/[\ion{Ne}{3}] by more than an order of magnitude, even with the lowest value of the ionization parameter, further suggesting that the hydrogen density is not constant. 
However, this confirms that the high density inferred in the central part of the \hii{} is not causing the elevated \ion{N}{3}] and \ion{N}{4}] emission compared to \ion{O}{3}] and \ion{C}{3}], especially since \ion{N}{3}] and \ion{C}{3}] are produced in the same zone of the \hii{} region.

Next, we explore whether depletion of oxygen and carbon onto dust grains could bring the predicted line ratios into agreement on its own.
We compute photoionization models with $\xi_d$ ranging from 0.1 to an extreme 0.95 ($\sim 3\times$ the solar value). 
The highest value of $\xi_d$ barely reach the observed \ion{N}{3}]/\ion{O}{3}] and \ion{N}{3}]/\ion{C}{3}] ratios, as both carbon and oxygen are depleted onto dust grains but not nitrogen. 
\ion{N}{3}]/H$\delta$ and \ion{N}{4}]/H$\delta$ are also increased due to the increase in electronic temperature and the absorption of H-ionizing photons. 
However, the flux in the neon lines provide a crucial check on this scenario since neon is also non-refractory.
At these high $\xi_d$ values necessary to reproduce \ion{N}{3}]/\ion{O}{3}] and \ion{N}{3}]/H$\delta$, the predicted [\ion{Ne}{3}]/[\ion{O}{2}] and [\ion{Ne}{3}]/\ion{C}{3}] are also strongly enhanced, and exceed the observed ratios by approximately an order of magnitude.
Thus, depletion does not successfully reproduce all line ratios simultaneously and cannot explain the apparent enhancement in the nitrogen lines.

With alternatives ruled-out, we predict the impact of significantly-enhanced nitrogen abundances on the UV--optical nebular lines.
We compute and show models with N/O enhanced by up to $(\mathrm{N/O})_{\mathrm{tot}}=0.7$ in the left panel of Figure~\ref{fig:photomod}.
This brings the modeled line ratios into good agreement with those observed, at super-solar nitrogen abundances in qualitative agreement with those estimated by \citetalias{cameronNitrogenEnhancements4402023}.

Leveraging the grid of photoionization models we have computed and the other line ratios available for \gnz{}, we proceed to directly fitting these lines with \texttt{BEAGLE} \citep{chevallardModellingInterpretingSpectral2016}.
We vary the metallicity within the range $[-1.5,0]$ in $\log_{10}(Z/Z_\odot)$, with the ISM metallicity set equal to the stellar metallicity, $\log_{10}(\Uav)$ varies within $[-4,-1]$, the age of the stellar population within $[5,9]$ in $\log_{10}(t/\mathrm{yr})$ assuming a constant star-formation history for the present modeled episode and the stellar mass within $[1,15]$ in $\log_{10}(M/M_\odot)$.
We adopt an SMC attenuation curve \citep{peiInterstellarDustMilky1992} and fit the V-band attenuation optical depth, \tauV, in the range 0 to 5. 
The total N/O is linked to total O/H through the relation :
\begin{equation}\label{eq:NOrel}
    \log_{10}(\rm{N/O})_{\rm{tot}} \approx \log_{10} \{ 10^{-1.732}+10^{[\rm{log_{10}(O/H)_{\rm{tot}}}+2.19]}\} + \rm{C}
\end{equation}
We fit C within the range $[-0.25,1.9]$. C=-0.25 corresponds to the N/O versus O/H relation of \citet{gutkinModellingNebularEmission2016}, which gives $\log_{10}(\mathrm{N/O})_{\mathrm{tot}}=-1.15$ at Z=Z$_\odot$ and C=1.9 gives $\log_{10}(\mathrm{N/O})_{\mathrm{tot}}=1$ at Z=Z$_\odot$. 
We fix $\xi_d$ to the solar value, 0.3, accounting for moderate dust depletion.
C/O abundance follows the relation between $\log_{10}(\rm{C/O})_{\rm{tot}}$ and $\rm{log_{10}(O/H)_{\rm{tot}}}$ derived by \citet{nichollsAbundanceScalingStars2017}. 
We include the following observables in the fit: \ion{N}{4}]~$\lambda 1485$, the upper limit of \ion{O}{3}]~$\lambda 1664$, \ion{N}{3}]~$\lambda 1750$, \ion{C}{3}]~$\lambda 1908$, [\ion{O}{2}]~$\lambda 3727$, [\ion{Ne}{3}]~$\lambda 3868$, [\ion{He}{1}]~$\lambda 3889$, [\ion{Ne}{3}]~$\lambda 3970$ + H$\epsilon$, H$\delta$, H$\gamma$ and [\ion{O}{3}]~$\lambda 4363$, as well as EW(H$\delta$) and EW(H$\gamma$). 

This fit results in a total metallicity estimate $\log_{10}(Z/Z_\odot)=-0.94_{-0.06}^{+0.06}$ and C=$1.14_{-0.03}^{+0.04}$.
This yields an estimated gas-phase abundances of $12+\log_{10}(\mathrm{O/H})_{\mathrm{gas}}=7.71$ and $\log_{10}(\mathrm{N/O})_{\mathrm{gas}}=-0.25$; and in-total (including the depleted material), $12+\log_{10}(\mathrm{O/H})_{\mathrm{tot}} = 7.84_{-0.05}^{+0.06}$ and $\log_{10}(\mathrm{N/O})_{\mathrm{tot}} = -0.38_{-0.04}^{+0.05}$ (substantially super-solar, at $[\mathrm{N/O}]=+0.52$; \citealt{caffauPhotosphericSolarOxygen2008}).
This N/O inference is significantly more tightly-constrained but broadly consistent with the lower end of the range estimated from a two-zone nebular model argument by \citet{cameronNitrogenEnhancements4402023}.
We also find a young best fit age of $6.57_{-0.2}^{+0.09}$ in $\log_{10}$(age/yr) corresponding to 3.7~Myr; and a large ionization parameter $\log_{10}(\Uav)=-1.63_{-0.08}^{+0.07}$.
These quantitative inferences from the nebular line spectrum add further support to the emerging picture of a galaxy dominated by extremely dense conglomerations of massive stars, with low overall metallicity despite a remarkably strong enhancement in nitrogen.

In sum, our exploration of a wide range of ionization mechanisms and gas conditions confirm that the observed UV--optical emission lines in \gnz{} require a substantially super-solar N/O; a conclusion we demonstrate is remarkably robust to assumptions about non-stellar ionizing agents, extremely high gas densities, and dust depletion of oxygen.
The \ion{N}{4}] emission is strong relative to \ion{N}{3}] but reproducible by a metal-poor ionizing spectrum under high-density conditions.
While we cannot definitively rule-out other ionizing sources, we do not find any distinct evidence for AGN or shocks among the nebular line constraints.

\section{Insight from local and lower-redshift samples}
\label{sec:localref}

The \ion{N}{4}] and \ion{N}{3}] transitions so prominent in \gnz{} are uncommonly detected in star-forming galaxies at high redshift.
However, they are not entirely without precedent among local metal-poor galaxy samples.
Comparisons to the local Universe can provide helpful reference points for understanding spectral properties in the distant Universe, even when exact matches cannot necessarily be found (see for instance the substantial insight gained on \ion{C}{3}] emission glimpsed in the reionization era pre-\jwst{}; e.g.\ \citealt{rigbyStarFormation4812017,jaskotPhotoionizationModelsSemiforbidden2016,senchynaUltravioletSpectraExtreme2017,senchynaExtremelyMetalpoorGalaxies2019}).

We perform a search for these emission lines in nearby star-forming galaxies observed in the ultraviolet.
A growing database of such spectra now exist especially from recent programs leveraging the sensitivity of \hstcos{}, targeting galaxies approaching the metallicities and stellar population ages likely to be encountered at high redshifts \citep[see e.g.\ the COS Legacy Archive Spectroscopic Survey, or CLASSY;][]{bergCOSLegacyArchive2022a,jamesCLASSYIITechnical2022}.

We first identify and examine \ion{N}{4}]-emitters in local star-forming samples in Section~\ref{sec:localn4}.
Then, we explore a unique galaxy in the local Universe which powers a remarkably similar UV spectrum in Section~\ref{sec:mrk996}.
Finally, we zoom back out in Section~\ref{sec:highzcomp} to briefly summarize and discuss galaxies with prominent UV nitrogen lines at other redshifts $z<10$.

\subsection{Local \ion{N}{4}]-emitters}
\label{sec:localn4}

\begin{figure}
    \centering
    \includegraphics[width=0.4\textwidth]{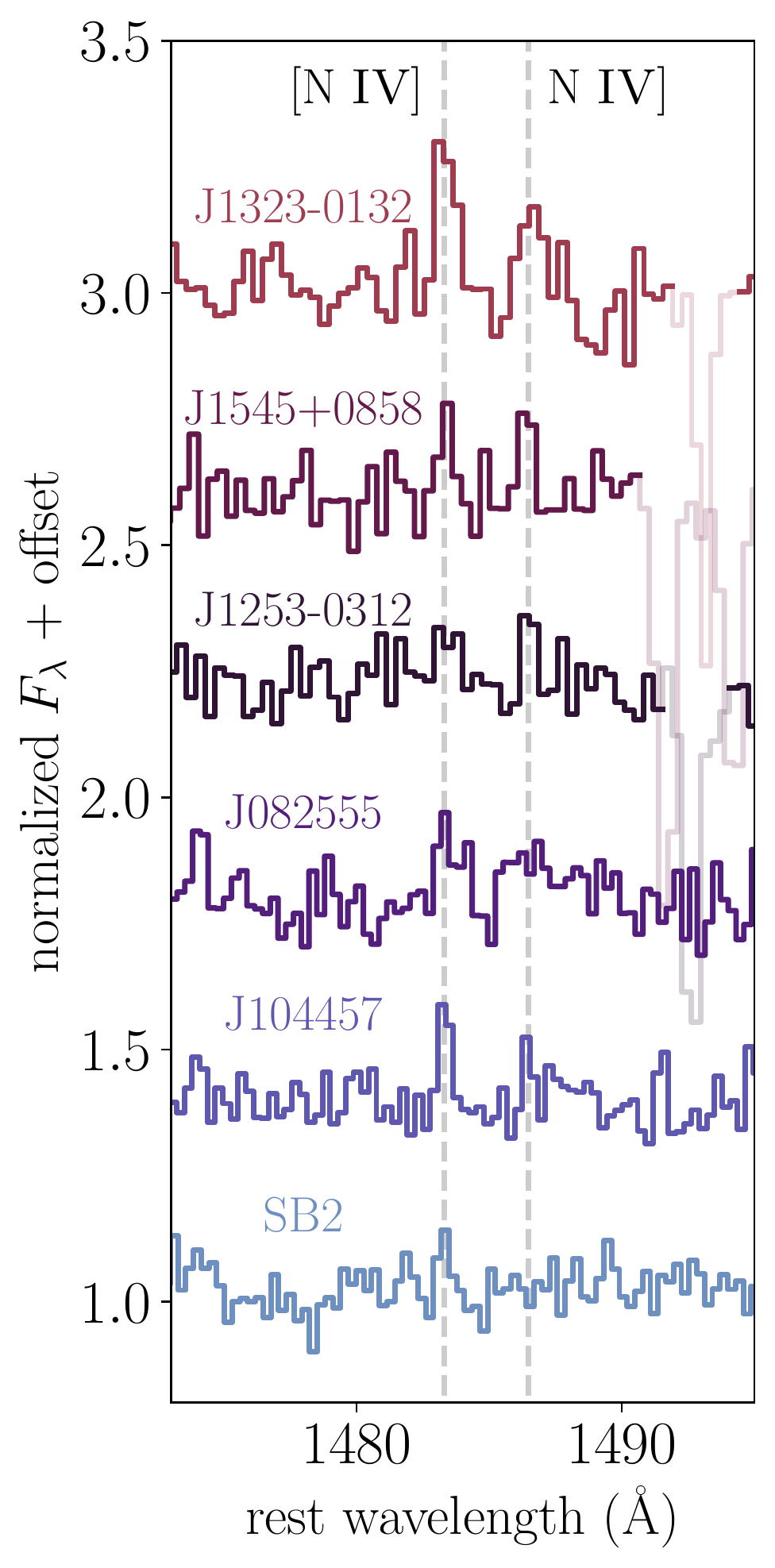}
    \caption{The \ion{N}{4}] doublet is encountered in deep UV spectra of local metal-poor star-forming galaxies, especially at low gas-phase metallicities $12+\log\mathrm{O/H}<8.0$ and young stellar ages.
    A selection of \hstcos{} detections of the doublet from the samples assembled by \citet{senchynaDirectConstraintsExtremely2022} and \citet{bergCOSLegacyArchive2022a} are plotted here, with MW and ISM absorption lines masked where present.
    The line ratios are consistent with more prosaic electron densities (range plotted in Figure~\ref{fig:niv_ratio}) than \gnz{}, and are substantially less prominent than in \gnz{}, consistent with the low nitrogen abundances in these local systems.
    }
    \label{fig:nivlocal}
\end{figure}

We search for \ion{N}{4}] detections in public COS UV spectra, focusing in particular on galaxies with medium-resolution grating data sufficient to resolve the \ion{N}{4}] components and ensure confident line identifications in the low signal-to-noise regime.
We find detections of \ion{N}{4}] $\lambda \lambda 1483, 1486$ emission in at least eight systems across the samples presented by \citet{bergCOSLegacyArchive2022a,senchynaDirectConstraintsExtremely2022}.
We plot the detections for a representative subset in Figure~\ref{fig:nivlocal}.

The galaxies powering nebular \ion{N}{4}] emission identified here tend to be metal-poor.
Of the six found in the CLASSY sample \citep[see also][]{mingozziCLASSYIVExploring2022}, all fall at $12+\log(\mathrm{O/H})\lesssim 8.0$ (approximately 20\% solar), and extend to the lowest-metallicities in that sample (e.g.\ J104457, at $12+\log\mathrm{O/H}=7.45$).
While limited by the sample size and the non-uniform depth of the parent samples of UV spectra, this suggests an empirical association between metal-poor massive stars and \ion{N}{4}] emission locally.

This association with low-metallicity stellar $Z/Z_\odot<0.2$ stellar populations is evocative of that identified in \ion{C}{4} emission \citep{senchynaExtremelyMetalpoorGalaxies2019,senchynaDirectConstraintsExtremely2022}.
Both \ion{N}{4} and \ion{C}{4} have very similar ionization potentials ($\sim 47$~eV), just below the $\mathrm{He}^+$-ionizing edge.
Indeed, four of the six CLASSY \ion{N}{4}]-emitters are also found to power \ion{C}{4} in nebular emission. 
To this sample, we add two additional \ion{N}{4}] and \ion{C}{4} emitters from the sample of \citet{senchynaDirectConstraintsExtremely2022}; for a detection rate of \ion{N}{4}] alongside \ion{C}{4} of half in this sample.
This suggests these lines share a common ionizing source and association with young metal-poor stellar populations; but that other factors, likely including both \ion{C}{4} resonant scattering \citep[e.g.][]{senchynaDirectConstraintsExtremely2022} and the N/O abundance and concentration of star formation (Section~\ref{sec:photomod}) modulate this association.

We also note for completeness detections of similar equivalent width \ion{N}{4}] $\lambda 1486$ with weak [\ion{N}{4}] $\lambda 1483$ in several of the Wolf-Rayet galaxies presented by \citet{senchynaUltravioletSpectraExtreme2021}.
At first glance this suggests an occurrence at much higher metallicity than just discussed, and potentially under density conditions approaching those in \gnz{}.
However, this line (though not \ion{N}{3}]) is also seen powered directly in Wolf-Rayet and similar stellar atmospheres themselves \citep[e.g.][]{martinsPropertiesWNhStars2009,hainichWolfRayetStarsLarge2014}.
And indeed, the detection of \ion{N}{4}~$\lambda 1719$ in P-Cygni in the same spectra at comparable strength confirm we are likely viewing wind emission in \ion{N}{4}].
We note that this is not a plausible explanation for the far more prominent emission in \gnz{}, where no clear stellar wind signatures are evidence (and in particular, comparable strength features in at least \ion{He}{2}, \ion{N}{5}, and \ion{N}{4} would be demanded).

While we cannot rule-out other ionizing contributions, the close association between \ion{C}{4}, \ion{N}{4}], and metal-poor star forming environments is suggestive of an association with the hard ionizing continuua powered by low-metallicity stellar populations locally.
Broadly, the line fluxes we derive are compatible with the photoionization model predictions for star-forming galaxies with nominal N/O presented in Section~\ref{sec:photomod}.
The detection statistics discussed above suggest that \ion{N}{4}] emission may be (like \ion{C}{4}) an indicator for young metal-poor stellar populations in the absence of non-stellar ionizing sources.

However, the doublet ratios and thus inferred densities in these local \ion{N}{4}]-emitters differ systematically from that measured in \gnz{}.
We fit this doublet ratio in the local blue compact dwarf spectra plotted in Figure~\ref{fig:nivlocal} following a similar procedure to that outlined in Section~\ref{sec:niv}, and find doublet ratios $F_{1483}/F_{1486}$ ranging from $0.95$ up to values consistent within the uncertainties with the high-density limit of $\sim 1.5$.
This suggests densities in the \ion{N}{4}]-emitting gas closer to $10^4$--$10^5$~$\mathrm{cm^{-3}}$ or below in the \ion{N}{4}]-emitting gas among these local galaxies (Figure~\ref{fig:niv_ratio}), as found by \citet{mingozziCLASSYIVExploring2022} in the CLASSY subset.

The cause of these systematically lower inferred nebular densities with respect to \gnz{} may reside in the significantly different global densities of star formation probed in these environments.
Even in the intense local \ion{C}{4}-emitters, dominated by young star formation in UV-compact regions, we are viewing populations characterized by $\Sigma_{\mathrm{SFR}}\lesssim 10\, \mathrm{M_\odot/yr/kpc^2}$ \citep[estimated from photometric parameters in][]{senchynaDirectConstraintsExtremely2022}.
This is $\gtrsim 4$ times below the unresolved lower limit placed upon \gnz{} (Section~\ref{sec:intro}), suggesting that the star formation mode that characterizes \gnz{} is substantially more concentrated and potentially highly-clustered than even these local blue compact dwarfs.

The other key difference with respect to \gnz{} is the relative strength of this \ion{N}{4}] emission.
In the local metal-poor star-forming systems we analyze here, we measure combined equivalent widths for the \ion{N}{4}] doublet ranging from $0.1$--$0.6$~\AA{}; substantially less prominent than the $\lambda 1486$ detection at $\sim 10$~\AA{} in \gnz{}
Unfortunately, we lack strong constraints on the \ion{N}{3}] line for these systems; none are detected, but the declining throughput of the COS/G160M grating at those longer wavelengths prevent us from ruling-out \ion{N}{3}] at a comparable $\lesssim 1$~\AA{} equivalent width to \ion{N}{4}].
But the reasonable agreement between the \ion{N}{4}] line fluxes relative to \ion{O}{3}] measured in these systems and predicted in Section~\ref{sec:photomod} without nitrogen enrichment strongly suggests that the root cause of this offset lies in their known lower nitrogen abundances \citep[e.g.][]{bergCharacterizingExtremeEmissionline2021}.

These local detection statistics indicate that \ion{N}{4}] is associated with young stellar populations at low metallicities ($Z/Z_\odot \lesssim 0.2$).
However, the doublet ratios and equivalent widths measured among these local systems differ significantly from those reported in \gnz{}.
Broadly, this lends further credence to the idea that \gnz{} is host to remarkably dense and highly nitrogen-enriched gas that is extremely rare even in galaxies dominated by recent metal-poor star formation nearby.

\subsection{A near-analogue for \gnz{}: Mrk~996}
\label{sec:mrk996}

\begin{figure}
    \centering
    \includegraphics[width=0.48\textwidth]{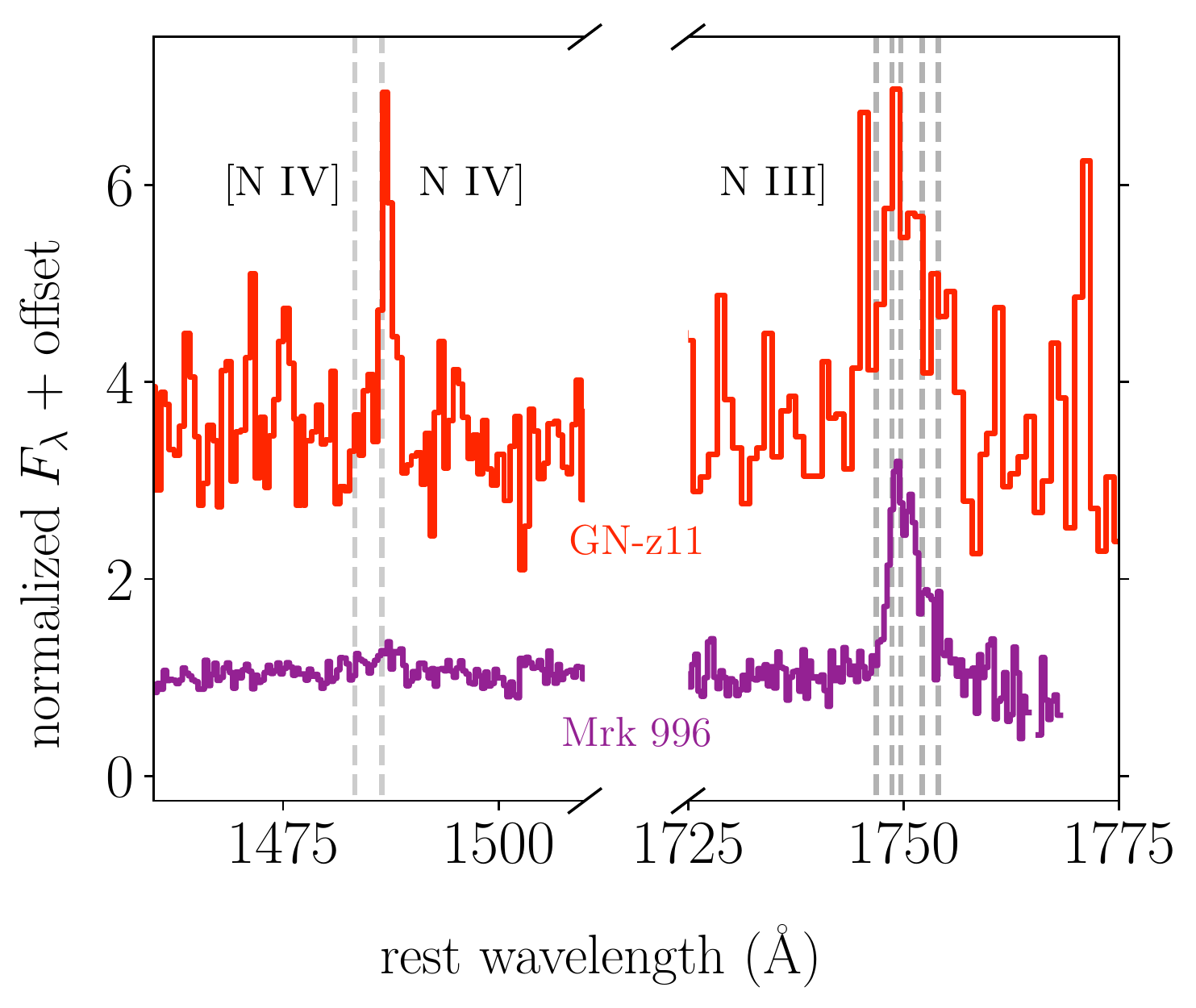}
    \caption{
        The \hstcos{} UV spectrum of the local blue compact dwarf galaxy Mrk~996 \citep{bergCOSLegacyArchive2022a} reveals \ion{N}{3}] emission strikingly similar to that observed in \gnz{}.
        A similar N/O overabundance and high central electron density is inferred in \mrk{}, and attributed to the presence of an unusual concentration of Wolf-Rayet stars with N-rich winds.
        This demonstrates that similar spectral features can be powered by massive stars embedded in their own dense CNO-processed ejecta.
        The addition of \ion{N}{4}] emission and lack of strong Wolf-Rayet features in \gnz{} further suggests that though similar, \gnz{} may be powered by lower-metallicity massive star populations.
    }
    \label{fig:mrk996comp}
\end{figure}

The local \ion{N}{4}-emitters discussed above are missing two key features that distinguish \gnz{}: the combination of very high $\sim 10^6$~$\mathrm{cm^{-3}}$ nebular electron densities and highly elevated N/O sufficient to boost the UV nitrogen lines.
However, a further review of the literature and local samples does reveal one nearby galaxy with attributes remarkably similar to those of \gnz{}.
The UV spectrum of \mrk{} stands out among the CLASSY sample of local star-forming galaxies with its extremely prominent \ion{N}{3}] emission complex; the only example therein of a clear detection of this complex \citep{mingozziCLASSYIVExploring2022}.
The equivalent width of this \ion{N}{3}] complex is 10~\AA{}; equal to that reported in \gnz{} by \citetalias{bunkerJADESNIRSpecSpectroscopy2023} (Figure~\ref{fig:mrk996comp}).
The scale of this star-forming complex is substantially smaller than \gnz{}, with a star formation rate an order of magnitude smaller \citep{jamesVLTVIMOSStudy2009}; however, a investigation of the conditions which support such similar \ion{N}{3}] emission is instructive.

While \ion{N}{3}] emission in the two systems is strikingly similar, \mrk{} does not power clearly detectable \ion{N}{4}] emission.
A hint of possible emission is evident in the $\lambda 1486$ component (Fig.~\ref{fig:mrk996comp}), but a comparable line EW is clearly ruled-out.
However, we have already demonstrated that the \ion{N}{4}]/\ion{N}{3}] ratio can vary by nearly 2 orders of magnitude with varying metallicity and ionization paramater over a reasonable range (Section~\ref{sec:photomod}).
Likewise, the detection rate of \ion{N}{4}] supports an association with ionizing radiation fields encountered at particularly low stellar metallicities (Section~\ref{sec:localn4})
The prominent \ion{N}{4}] $\lambda 1486$ accompanying \ion{N}{3}] in \gnz{} then could be attributed to a higher concentration of potentially more metal-poor ionizing stars than \mrk{}.

Like \gnz{}, \mrk{} is a clear outlier in its prominent \ion{N}{3}] emission from other star-forming galaxies which might otherwise appear qualitatively similar.
An explanation for this in \mrk{} is clearly revealed by multiwavelength studies.
The massive stars in \mrk{} are unusually-concentrated for a blue compact dwarf even in space-based optical imaging, with the star formation occuring almost entirely within an unresolved $r\lesssim 160$~pc central region \citep{thuanHubbleSpaceTelescope1996}.
While the lack of \ion{N}{4}] emission prevents a similar $n_e$ measurement, a combination of an anomalously high H$\alpha$/H$\beta$ ratio, peculiar \ion{O}{3} and \ion{He}{1} line ratios, and measurements of two density-sensitive doublets of [\ion{Fe}{3}] in the optical all suggest core central densities of $n_e\simeq 10^6$~$\mathrm{cm^{-3}}$ or slightly higher \citep{thuanHubbleSpaceTelescope1996,jamesVLTVIMOSStudy2009,tellesGeminiGMOSStudy2014b} -- entirely compatible with the high density we infer in the highly-ionized gas of \gnz{}.
The core region is clearly nitrogen-enhanced relative to the surrounding gas (for which they estimate $\log\mathrm{N/O}=-1.5$): with a super-solar $\log\mathrm{N/O}=-0.15$ inferred by \citet[][]{jamesVLTVIMOSStudy2009,tellesGeminiGMOSStudy2014b}.
This extreme overabundance is also in good agreement with that inferred from the UV nitrogen and oxygen lines in \gnz{}.
The oxygen abundance in the core is less clear; the surrounding gas appears to be metal-poor, with $12+\log\mathrm{O/H}=7.9$ \citep{tellesGeminiGMOSStudy2014b}; but \citet{jamesVLTVIMOSStudy2009} estimate a higher central O/H closer to half-solar.
A somewhat higher stellar metallicity $Z/Z_\odot\gtrsim 0.2$ is suggested by the prominence of stellar wind features in the UV--optical, and again the correspondingly softer radiation field would also naturally explain the lack of \ion{N}{4}] emission.

In any case, the picture assembled in \mrk{} also points to a clear origin of the extremely dense, nitrogen-enriched gas at the core of the cluster.
The signatures of Wolf-Rayet winds are clear in the optical spectra, entirely concentrated in the dense nuclear cluster.
Even without correcting the intrinsic wind line luminosities for the potentially low stellar metallicity of \mrk{}, \citet{tellesGeminiGMOSStudy2014b} estimate an extremely high ratio of $N(\mathrm{WR})/N(\mathrm{O+WR}) = 0.19$.
This association strongly suggests that we are observing the highly nitrogen-enriched ejecta of these massive stars.
Signatures of broadening in some of the optical lines suggest an isotropic outflow forming in the center, plausibly launched by the mechanical energy input by these WR winds.

\mrk{} provides a demonstration that photoionization of highly N-enriched stellar ejecta in a dense stellar cluster can produce UV spectral features similar to those glimpsed in \gnz{}.
Unfortunately, \mrk{} is essentially entirely unique among blue compact dwarfs --- no other star-forming galaxy with such prominent localized Wolf-Rayet enrichment or a similar UV spectrum has been identified in the local Universe.
Given that the peculiar N/O enhancement and electron density of \mrk{} can be identified from fairly shallow optical spectroscopy and that the SDSS and other spectroscopic surveys have not identified any similar objects \citep[e.g.][]{pilyuginAbundanceDeterminationGlobal2012}, \mrk{} appears to be the exception that proves the rule: the peculiar set of processes occuring in \gnz{} are likely exceptionally rare to absent at $z\sim 0$.

\subsection{Other high-$z$ reference points}
\label{sec:highzcomp}

Before returning to focus on \gnz{} in the following discussion section, we briefly summarize here other galaxies beyond the nearby $\lesssim 200$~Mpc Universe with reported \ion{N}{3}] or \ion{N}{4}] emission for comparison.

First, a super star cluster complex in the Sunburst Arc stands out as among the most instructive comparison points.
We return to discuss the implications of a detailed study of this system \citep{pascaleLymancontinuumleakingSuperStar2023} in Section~\ref{sec:summary}, but note here that it powers an \ion{N}{3}] complex nearly as strong relative to the other UV lines as observed in \gnz{}.

These transitions have been noted in several other apparently star-forming galaxies at intermediate to high redshifts.
The Lynx arc powers a spectrum with prominent \ion{N}{4}] alongside similarly-strong \ion{C}{4} and weaker \ion{N}{3}] \citep{fosburyMassiveStarFormation2003}, which they argue are powered by star formation; notably, the \ion{N}{4}] ratio is not in the high-density limit as observed for \gnz{}.
At higher redshift, \citet{raiterLyaEmittersGOODSS2010} and \citet{vanzellaUnusualIVEmitter2010} identify a spectrum dominated by \ion{N}{4}] apparently in the high-density limit in a $z=5.56$ Ly$\alpha$-emitter with no other rest-UV metal lines; the relative role of very metal-poor star formation and a possible low-luminosity AGN in shaping this spectrum remains unclear.
\citet{mcgreerBrightlensedGalaxyStrong2018} report the detection of \ion{N}{4}] in the high-density limit and possible nebular \ion{C}{4} in a lensed $z=5.42$ system which they argue is best reproduced by star formation.
This systems is the closest in SFR and mass scale to \gnz{} among these systems; though uncertain due to lensing, it is plausibly forming stars at $10$--$100\,\mathrm{M_\odot/yr}$.

Finally, \ion{N}{4}] emission is also detected among some AGN.
\citet{hainlineRestframeUltravioletSpectra2011} find a detection of \ion{N}{4}] at lower flux than \ion{N}{5} and \ion{C}{4} in a stack of UV-selected AGN at $z\sim 2$--3;
The strength of \ion{N}{4}] in a sample of $z\sim 2.5$ radio galaxies leads \citet{humphreyDeepSpectroscopyFUVoptical2008} to invoke a combination of AGN and shock ionization to explain their presence.
\ion{N}{4}] emission is not entirely uncommon in deep spectra of QSOs, and has been leveraged even as a metallicity indicator \citep[e.g.][]{hamannElementalAbundancesQuasistellar1999}.
However, a class of rare quasar has been defined by the prominence of broad emission in \ion{N}{4}], \ion{N}{3}], and \ion{N}{5}: the `nitrogen-loud' quasars \citep[e.g.][]{glikmanDiscoveryTwoSpectroscopically2007,batraMetallicitiesBroadEmission2014,matsuokaChemicalEnrichmentAccretion2017}.
Interestingly, \citet{matsuokaChemicalEnrichmentAccretion2017} argue that the spectra of these objects likely reflect substantial nitrogen enrichment, plausibly related to intense nuclear star formation.

\section{Discussion and summary: \gnz{} as a potential globular cluster precursor}
\label{sec:summary}

Our re-appraisal of the peculiar rest-UV spectrum of \gnz{} reveals several key clues about the system.
Regardless of ionization mechanism, the \ion{N}{4}] and \ion{N}{3}] emission in \gnz{} cannot be reproduced without the presence of very dense ($n_e\gtrsim 10^5$~$\mathrm{cm^{-3}}$; Section~\ref{sec:niv}) and highly nitrogen-enriched ($\log_{10} \mathrm{N/O}= -0.38$; Section~\ref{sec:photomod}) gas.
A close comparison to the local compendium of UV star-forming galaxy spectra suggests that the relative strength of \ion{N}{4}] in particular is consistent with ionizing spectra of particularly low-metallicity ($Z/Z_\odot<0.2$) and very young stellar populations (Section~\ref{sec:localref}).
However, the densities inferred from \ion{N}{4}] in these local systems are systematically lower than that implied in \gnz{}, suggesting that \gnz{} might be undergoing a far more dense mode of star formation that is extremely rare today.
Close similarities with a unique high-density star-forming complex in \mrk{} are noted, which support the idea that the peculiar properties of \gnz{} may be related to conditions of extremely high-density massive star formation.
A natural question that then arises is whether any record of such an event might be recorded in the local Universe.

The nitrogen enrichment is a crucial clue in contextualing \gnz{}.
The major site of nitrogen production in the Universe is in massive stellar interiors fusing hydrogen into helium following the CNO process.
This process utilizes carbon, nitrogen, and oxygen isotopes as catalysts, and is bottlenecked by the slow proton capture step $^{14}\mathrm{N}(\mathrm{p},\gamma)^{15}\mathrm{O}$; thus efficiently converting C and O to N as a byproduct of hydrogen fusion.
In contrast to the $\alpha$-elements carbon and oxygen, then, substantial nitrogen is not naturally expected as a generic product of any class of supernovae, including primordial Population III models \citep[as explored by ][]{cameronNitrogenEnhancements4402023}.
Instead, the ISM is enriched with nitrogen primarily through processes which can eject CNO-processed material from stellar interiors during their lifetime --- including the winds of Wolf-Rayet and similar stars, where the N-rich nuclear-processed core layers of a massive star have been exposed through wind or binary mass stripping.
Indeed, galaxies in the local Universe presently hosting large populations of Wolf-Rayet stars commonly show enhancements in N/O, though generally of much smaller scale than in \mrk{} \citep[e.g.][]{pagelNewMeasurementsHelium1986,brinchmannGalaxiesWolfRayetSignatures2008}.

\begin{figure}
    \centering
    \includegraphics[width=0.48\textwidth]{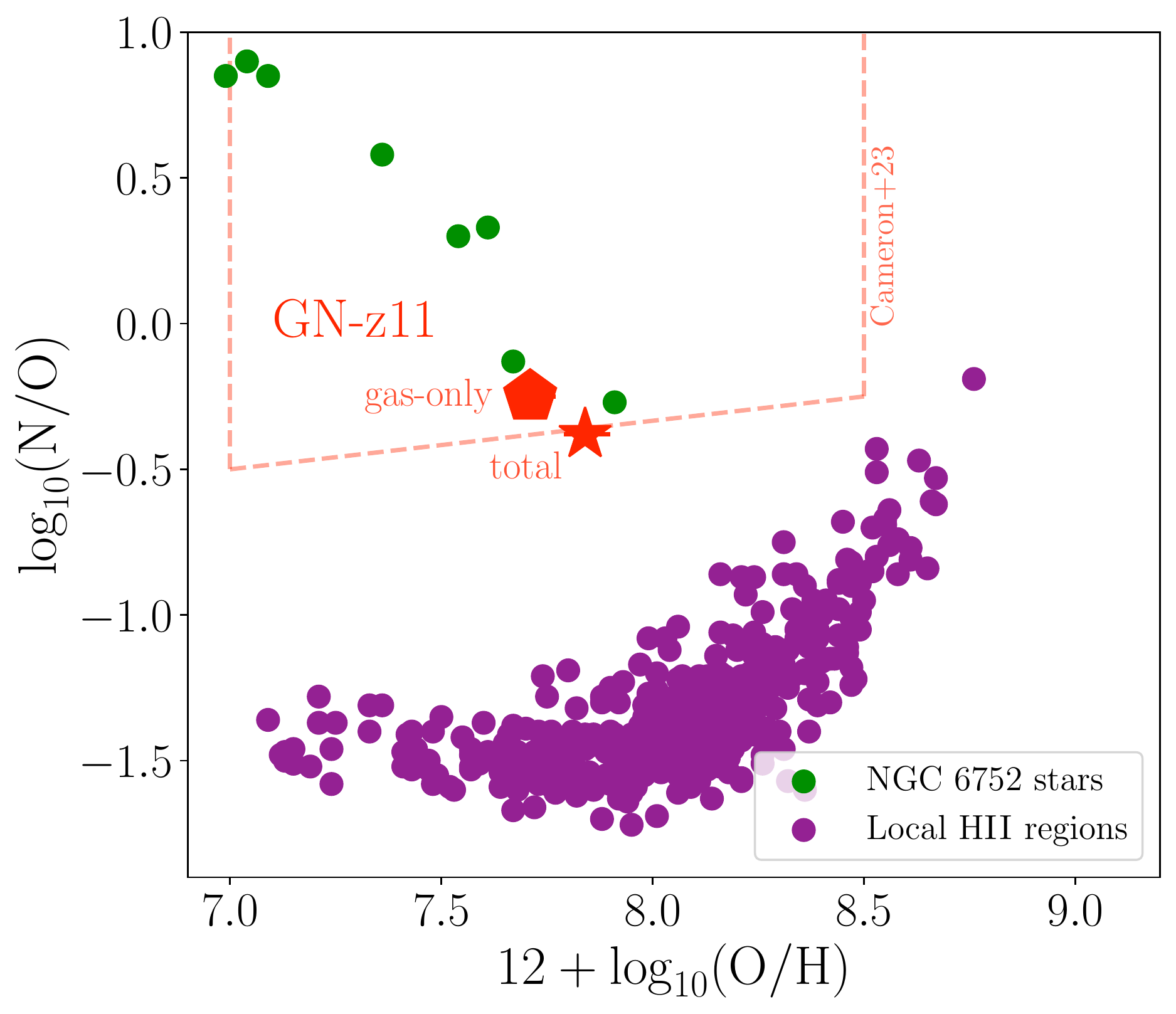}
    \caption{
        The O/H-N/O abundance space occupied by \gnz{} is in good agreement with the patterns observed in present-day globular clusters.
        To illustrate, we compare constraints on \gnz{} from both \citet{cameronNitrogenEnhancements4402023} (red box) and our photoionization modeling results (red symbols, errorbars partially covered; Section~\ref{sec:photomod}) to abundances of dwarf turnoff stars in the globular cluster NGC~6752 \citep{carrettaAbundancesSlightlyEvolved2005}. 
        We also plot the abundances of local \hii{} regions sampled by \citet{pilyuginCounterpartMethodAbundance2012}.
        The location of \gnz{} close to the locus of globular cluster stars at systematically lower O/H and higher N/O relative to local star-forming systems suggests that we may be viewing (with potentially some dilution by other star formation) the dense N-rich ejecta from early stellar populations embedded in globular cluster precursors in \gnz{}.
    }
    \label{fig:abundcomp}
\end{figure}

In the local Universe, metal-poor stellar populations with evidence for initial abundances enriched by CNO-processing are rare in the field and open clusters \citep[e.g.][]{magriniGaiaESOSurveyAbundance2018}, but they are a defining feature of globular clusters \citep[e.g.][]{grattonAbundanceVariationsGlobular2004}.
If stars form or are currently forming from ISM material in \gnz{} with the abundance pattern we are glimpsing in the ionized gas phase, they will closely resemble in N and O the enriched populations we see today in nearby globular clusters.
To illustrate this, we compare the gas-phase and total N/O we infer in \gnz{} with measurements reported for stars in the nearby globular cluster NGC~6752 by \citet{carrettaAbundancesSlightlyEvolved2005} in Figure~\ref{fig:abundcomp}.
In particular, we plot the abundances derived for faint dwarf main sequence turn-off stars, where no mixing or dredge-up is expected to modulate photospheric CNO abundances \citep[see also e.g.][]{brileyChemicalInhomogeneityFaint2004,brileyCarbonNitrogenAbundances2004}.
The overlap between these stars and the abundance estimates derived for \gnz{} is striking, in contrast to the locus of local \hii{} regions from \citet{pilyuginCounterpartMethodAbundance2012} with which the vast majority of star-forming galaxies at all redshifts overlap \citep[see also][]{bergCHAOSIVGasPhase2020,cameronNitrogenEnhancements4402023}.

This interpretation has significant potential implications for the origin of globular clusters.
First, our results clearly establish the basic plausibility of the natal enrichment picture for the second generation in the most ancient globulars, as we are viewing in-situ a quantity of N/O-enhanced material more than sufficient to form a globular cluster population.
To estimate the total mass of material involved, we perform order-of-magnitude calculations in Appendix~\ref{app:nitrogen} that establish that the mass of gas in the ionized phase with integrated abundances reflecting CNO-processing easily exceeds $10^7$~$\mathrm{M_\odot}$.
Even without accounting for enriched material in other phases and allowing for an extremely low star formation conversion efficiency, this easily represents enough material to form massive globulars.

Indeed, the scale of this mode of star formation is strikingly large and early.
Relying on stellar mass estimates from photometric fitting \citep{tacchellaJADESImagingGNz112023}, we may be viewing upwards of $10^8$~$\mathrm{M_\odot}$ of stars built-up already in what appears to be a system currently dominated by globular-like gas abundances, packed into an extremely compact $<200$~pc region.
Comparing the observed intrinsic luminosity of \gnz{} \citep[$M_{\mathrm{UV}}=-21.6$;][]{tacchellaJADESImagingGNz112023} to model predictions of very young $<5.5$~Myr globulars in-formation from \citet{boylan-kolchinLittleEnginesThat2018} suggests \gnz{} is an order of magnitude more luminous than expected for the brightest individual globulars based upon their present-day stellar mass.

It is instructive to compare \gnz{} to another well-studied high-redshift system, but one whose detailed investigation is aided by its lower redshift and strong gravitational lensing: the Sunburst Arc at $z=2.4$.
The Sunburst is host to a compact young super star cluster \citep[$\sim 3$~Myr;][]{chisholmConstrainingMetallicitiesAges2019} which is leaking Lyman continuum radiation \citep{rivera-thorsenGravitationalLensingReveals2019}, and whose UV spectrum bears some similarity to \gnz{} and \mrk{}.
In particular, strong \ion{N}{3}] $\lambda 1750$ emission is apparent, though at a somewhat lower flux relative to \ion{O}{3}] than in these systems (\ion{N}{3}] $\lambda 1750$/\ion{O}{3}] $\lambda\lambda 1661,1666$ near 0.9, c.f. $>2$ in \gnz{} and \mrk{}; Figure~\ref{fig:photomod} and \citealt{pascaleLymancontinuumleakingSuperStar2023}).
\citet{pascaleLymancontinuumleakingSuperStar2023} conclude from a detailed analysis (leveraging strong constraints on the hydrogen column density and cluster radius unavailable for \gnz{}) that this cluster is host to $\gtrsim 10^5$~$\mathrm{M_\odot}$ of dense, highly nitrogen enriched (inferring $\log_{10}(\mathrm{N/O}) = -0.2$ comparable to our gas-phase measurement in \gnz{}) material which is likely capable of forming a globular cluster.
The similarity of \gnz{} to spectra of the Sunburst cluster (and \mrk{}) is particularly striking given that \gnz{} is at least an order of magnitude more massive, in both photometrically estimated stellar mass and inferred quantity of nitrogen-enriched gas involved.

It is likely that there is at least some UV and potentially nebular contribution from other star formation in \gnz{}.
Indeed, the position of the total abundances we infer in this work with respect to the range of globular cluster star abundances in Figure~\ref{fig:abundcomp} suggests that we could plausibly be analyzing nebular emission from globular precursors diluted by star formation proceeding in gas polluted with more typical N/O.
Even with some dilution, however, the brightness and mass scales at play suggest we may be viewing the simultaneous assembly of a system of globulars.
However, the actual number of systems likely to be involved is unclear, as is the stage of their formation.
It is possible that a small number of initially very massive clusters may have been caught during formation, after which substantial stellar mass may be lost to dynamical stripping and other evaporation mechanisms, as has been suggested as a solution to certain abundance modeling issues in globulars \citep[e.g.][]{decressinOriginAbundancePatterns2007}.
Regardless, our results strongly suggest that we are viewing a system within $\sim 400$~Myr of the Big Bang whose UV and ionizing light is dominated by stars forming under intense globular cluster-like conditions.

The exact origin of the nuclear-processed material which enriched present-day globular clusters has long been the focus of debate \citep[e.g.][]{bastianMultipleStellarPopulations2018}.
One of the long-standing candidates are the winds of intermediate-mass $\sim4$--$7$~$\mathrm{M_\odot}$ asymptotic giant branch (AGB) stars \citep[e.g.][]{venturaMassiveAGBModels2009}.
However, the characteristic timescales for such relatively low-mass stars to evolve off the main sequence and traverse the AGB phase are $\sim 40$--100~Myr.
As pointed out by \citet{cameronNitrogenEnhancements4402023}, this is at-odds with the apparently rising star formation history of \gnz{} situated at just $\sim 400$~Myr after the Big Bang and its uncertain but young light-weighted age estimate of $<20$~Myr.
While we cannot rule such sources out, capturing already extremely elevated N/O associated with a very young burst of star formation is suggestive of a prompt enrichment mechanism.

However, the spectrum of \gnz{} also casts doubt on another candidate mechanism for globular enrichment.
The winds of very massive stars potentially formed by stellar collisions and have also been invoked to produce the observed abundance patterns \citep[e.g.][also discussed by \citealt{cameronNitrogenEnhancements4402023}]{gielesConcurrentFormationSupermassive2018}.
However, the dense winds powered by such a massive star would be expected to produce prominent signatures in UV lines, especially resonant \ion{N}{5} $\lambda\lambda 1238,1240$ and \ion{C}{4} $\lambda\lambda 1548,1550$ but also lines like \ion{He}{2} $\lambda 1640$ which are enhanced in massive stars near the Eddington limit even at lower metallicities \citep[e.g.][]{senchynaUltravioletSpectraExtreme2021,martinsSpectroscopicEvolutionVery2022}.
While the \jwst{} spectra are limited in SNR and resolution, very strong P-Cygni profiles in these transitions appear ruled-out by the prism data, whose overall morphology is more consistent with the relatively blank continuum observed at stellar metallicities $Z/Z_\odot<0.2$ where metal line-driven winds are substantially weakened \citep[e.g.][]{senchynaUltravioletSpectraExtreme2017}; which is consistent with the low stellar metallicity suggested by both the gas-phase abundances and \ion{N}{4}] detection.

The combination of apparently prompt enrichment without luminous massive star wind features instead points us to other mechanisms by which CNO-processed material might be ejected by massive stars.
Even in present-day open clusters, the majority of massive stars are not just in binaries but are expected to interact and undergo mass transfer or even merge with a companion \citep[e.g.][]{sanaBinaryInteractionDominates2012}.
As explored by \citet{deminkMassiveBinariesSource2009}, this mass transfer is likely to often be highly non-conservative and eject substantial amounts of nuclear-processed material to the cluster medium, readily reproducing globular cluster abundance patterns in broad strokes.
This material has the advantage of being ejected at much slower velocities than typical radiatively-driven stellar winds, enhancing the likelihood of it staying bound to the cluster.
Additionally, this mechanism readily explains the enhancement in these yields in extremely dense early-Universe star formation conditions, as dynamical interactions with both cluster gas and other stars will form and harden close binaries and enhance the likelihood and rate of interactions extremely efficiently \citep[in addition to other processes with similar products such as stellar collisions; e.g.][]{fabryckyShrinkingBinaryPlanetary2007,deminkMassiveBinariesSource2009,sillsMultiplePopulationsGlobular2010,deminkIncidenceStellarMergers2014}.

While only limited conclusions can be drawn from a single object, it is possible that unlike \mrk{} nearby, \gnz{} may not be entirely unique in the high-redshift Universe.
While their nature likely varies, some other high-redshift detections of prominent \ion{N}{3}] and \ion{N}{4}] emission lines beyond \gnz{} and the Sunburst Arc (Section~\ref{sec:highzcomp}) may hint at similar conditions.
If globulars are a major source of UV light at $z>6$ \citep[e.g.][]{boylan-kolchinLittleEnginesThat2018}, it is likely that other precursor systems will be caught at similar phases to \gnz{} especially among UV-selected samples, enabling more detailed investigation of the timescales and processes involved in building-up their stellar mass and peculiar abundances.
Statistics of deep UV spectra of larger samples of systems to be collected with \jwst{} promise to provide powerful direct insight onto the formation conditions of globulars and other primordial structures complementary to the picture assembled from local stellar archaeology.

\begin{acknowledgments}

The authors acknowledge helpful conversations with Andrew McWilliam, Josh Simon, and Ylva G\"{o}tberg during the drafting of this manuscript.
P.~S.\ was generously supported by a Carnegie Fellowship through the Carnegie Observatories while conducting this work.

This research made use of Astropy, a community-developed core python package for Astronomy \citep{astropycollaborationAstropyCommunityPython2013}; Matplotlib \citep{hunterMatplotlib2DGraphics2007}; Numpy and SciPy \citep{harrisArrayProgrammingNumPy2020,virtanenSciPyFundamentalAlgorithms2020a}; the SIMBAD database, operated at CDS, Strasbourg, France; and NASA's Astrophysics Data System.
   
\end{acknowledgments}

\bibliography{zoterolib}{}
\bibliographystyle{aasjournal}

\appendix

\section{On the excess quantity of nitrogen}
\label{app:nitrogen}

A simplified order-of-magnitude calculation can provide guidance as to the amount of nitrogen enrichment necessary to explain the N/O enhancement in \gnz{}.
We follow the same approach as \citet{brinchmannNewInsightsStellar2008} to estimate a total mass of `excess' nitrogen in the system.
First, we calculate the total mass of ionized hydrogen under a simple Case B Str\"{o}mgren sphere via
\begin{displaymath}
    \mathrm{M}_{\mathrm{ionized \, H}} 
    = \frac{m_p L_{\mathrm{H\beta}}}{(1.238\e{-25}\, \mathrm{cm^3 \,erg\, s^{-1}})\, T_4^{-0.874}\, n_e}
\end{displaymath}
where $m_p$ is the proton mass, $T_4=T_e/10^4\, \mathrm{K}$, and $T_e$ and $n_e$ are the electron temperature and density \citep{drainePhysicsInterstellarIntergalactic2011a}.
For \gnz{}, we calculate $L_{\mathrm{H\beta}}$ by first adopting the H$\gamma$ flux in the grating spectrum reported by \citetalias[][]{bunkerJADESNIRSpecSpectroscopy2023}: $13.9\e{-19} \mathrm{erg s^{-1} cm^{-2}}$.
Converting to H$\beta$ luminosity per the Planck 2018 cosmology \citep{collaborationPlanck2018Results2020} and following the fiducial Case B recombination spectrum at $T_e=10^4\, \mathrm{K}$, $n_e=10^3\, \mathrm{cm^{-3}}$ \citep{drainePhysicsInterstellarIntergalactic2011a}, we adopt $L_{\mathrm{H\beta}} = 4.6\e{42} \,\mathrm{erg/s}$.
Adopting these same assumed $n_e$ and $T_e$, we estimate a corresponding mass of ionized hydrogen of $3\e{7}\, \mathrm{M_\odot}$.
Then defining $\Delta(\mathrm{N/H})$ as the excess in the nitrogen (number) abundance over that expected for the galaxy, we can write the excess mass in nitrogen as
\begin{displaymath}
    \Delta M_\mathrm{N} = \frac{m_\mathrm{N}}{m_\mathrm{H}} \cdot M_{\mathrm{ionized\, H}} \cdot \Delta \left(\mathrm{\frac{N}{H}}\right)
\end{displaymath}
where $m_\mathrm{N}/m_\mathrm{H} \simeq 14$ represents the average ratio of the masses of the nitrogen and hydrogen ions.

At the gas-phase oxygen abundance of $12+\log_{10}(\mathrm{O/H})_{\mathrm{gas}}= 7.7$ estimated in Section~\ref{sec:photomod}, the expected N/H based upon the locus of local SDSS galaxies is $\log_{10}(\mathrm{N/O})=-1.5$ and thereby $\log_{10}(\mathrm{N/H}) = -5.8$
The required boost in the gas-phase nitrogen to reach our inferred $\log_{10}(\mathrm{N/O})_{\mathrm{gas}} = -0.25$ leads us to $\log_{10}(\mathrm{N/H}) > -4.6$.
Thus, we require a $\Delta(\mathrm{N/H}) \simeq 2.4\e{-5}$ in the gas-phase.
Plugging this into the above, we thus find a total mass of nitrogen required of $1\e{4}$~$\mathrm{M_\odot}$.

As a rough check on the plausibility of this value, we compare it to an order of magnitude calculation of the number of massive stars present in the galaxy.
First, assuming that the Balmer line luminosity is dominated by O-stars and assuming $L(\mathrm{H\beta})=5\e{36}\,\mathrm{erg/s}$ for an O7V star, we very approximately estimate a total number of $10^6$ O-stars.
Following the canonical line of reasoning adopting a large but plausible binary-influenced WR/O ratio at this metallicity of $0.05$ \citep{eldridgeBinaryPopulationSpectral2017}, we thus estimate a mass of nitrogen per WR star of 0.2~$\mathrm{M_\odot}$.
This is large, but encompassed by the range of values inferred in a similar manner in SDSS galaxies by \citet{brinchmannGalaxiesWolfRayetSignatures2008}; and within an order of magnitude of the total mass of nitrogen ejecta implied by observations of Galactic WR ring nebulae \citep{estebanSpatiallyResolvedSpectroscopy1992}.
Alternatively, it is not unreasonable to suggest that with an open cluster interacting binary fraction of $>70\%$ \citep[e.g.][]{sanaBinaryInteractionDominates2012}, and dynamical interactions in a dense cluster potentially elevating both this fraction and the incidence of extreme interactions ejecting CNO-processed material \citep[e.g.][]{deminkMassiveBinariesSource2009}, a relatively prosaic nitrogen mass per `interaction' as low as $< 0.1$~$\mathrm{M_\odot}$ may be sufficient.

\label{lastpage}

\end{document}